\documentclass[12pt]{iopart}
 
\usepackage{mathrsfs}
\usepackage{amsfonts}
\usepackage{amsthm,amssymb}
\usepackage{lscape}
\usepackage{subeqn} 
\usepackage{iopams}  
\usepackage{graphicx}
\usepackage{graphics}
\usepackage{color}

\theoremstyle{plain}
\newtheorem{theorem}{Theorem}[section]
\newtheorem{proposition}{Proposition}[section]
\newtheorem{lemma}{Lemma}[section]
\newtheorem{corollary}{Corollary}[section]
\theoremstyle{remark}
\newtheorem{comments}{Comment}

\newcommand{\be}{\begin{equation}}
\newcommand{\ee}{\end{equation}}

\newcommand{\real}{\mathbb{R}}

\newcommand{\pd}[2]{\frac{\partial #1}{\partial #2}}

\newcommand{\hor}{{\cal{H}}}

\usepackage{graphics,epsf}
\usepackage{tikz}
\usepackage{amsfonts}
\usetikzlibrary{decorations.pathmorphing,patterns,calc}
\begin{document}

\newcommand{\mcv}{$\textrm{McV}{\!}_4\,$}


\title[Falling into a McVittie Black Hole]{Can you fall into a McVittie black hole? Will you survive?}
\author{Brien C. Nolan}

\address{Centre for Astrophysics and Relativity, School of Mathematical Sciences, Dublin City University, Glasnevin, Dublin 9, Ireland.}
\eads{\mailto{brien.nolan@dcu.ie}}

\begin{abstract}
Yes and maybe. In contrast to the fluid particles of this perfect fluid spacetime which follow non-geodesic world-lines and escape to infinity, we prove that freely-falling test particles of McVittie spacetime can reach the black hole horizon in finite proper time. We review the relevant evidence and argue that the fate of an extended test body is less clear. More precisely:  
we consider expanding McVittie spacetimes with a non-negative cosmological constant. In each member of this class, we identify a region of the spacetime such that an observer following an initially-ingoing timelike geodesic crosses the black hole horizon of the spacetime in a finite amount of proper time. The curvature behaves in interesting ways along these geodesics. In the case of a positive cosmological constant, curvature scalars (of zero, first and second order in derivatives), Jacobi fields and parallel propagated (p.p.) frame components of the curvature remain finite along timelike geodesics running into the black hole horizon. For a vanishing cosmological constant, scalar curvature terms of zero and first order as well as Jacobi fields remain finite in this limit. However, scalar curvature terms of second order diverge, and we show that there are p.p.\ frame components of the curvature tensor that also diverge in this limit. We argue that this casts a doubt as to whether or not an extended test body can survive crossing the black hole horizon in this case.  \end{abstract}

\maketitle


\section{Introduction}
A McVittie spacetime is a solution of the Einstein field equations with perfect fluid source that embeds the Schwarzschild field in a Friedmann-Lema\^itre-Robertson-Walker (FLRW) background \cite{mcvittie1933mass}. An obvious question arises as to whether or not this spherically symmetric spacetime models a black hole in an expanding universe: this is a non-trivial question about the global structure of the spacetime. The first study of the global structure was undertaken by Sussman \cite{sussman1988spherically}, and subsequent work demonstrated that McVittie spacetimes can indeed provide models of black holes embedded in an expanding isotropic background - see \cite{kaloper2010mcvittie} and \cite{lake2011more,nolan2017local}. To be more precise: in an expanding McVittie spacetime, all future-directed outgoing radial null geodesics are future-complete, and the area radius $r$ and cosmic time function $t$ both extend to infinity along these geodesics. However, there are future-directed ingoing radial null geodesics that meet, at finite affine distance, a future boundary of the past of future null infinity. This is the event horizon of the McVittie black hole (we will refer to it throughout at the black hole horizon). The area radius has a finite positive value at the black hole horizon, but the cosmic time function $t\to+\infty$ along ingoing radial null geodesics extending to the horizon. A more precise statement is given below (Proposition 1.1). 

On the other hand, the fluid particles of the spacetime do not fall into the black hole: $r$ diverges to positive infinity as $t\to+\infty$ along each fluid world line (see \cite{lake2011more} for a special case, and below for the general case). So can (other) material particles fall into a McVittie black hole? And if so, do they encounter divergent matter or curvature terms on crossing the horizon? There is reason to suspect that they do so: in fact it has been claimed recently that curvature scalars and matter invariants diverge at the event horizon of a McVittie black hole, and that therefore, the spacetime cannot provide a model of a physical black hole in an expanding universe \cite{Poplawski_2025}. However, this claim was made without analysing the behaviour of invariants along geodesics reaching the horizon, and indeed the issue has been addressed previously. In \cite{kaloper2010mcvittie}, it was shown that the relevant invariants are finite along ingoing radial null geodesics that meet the horizon: these results were extended and refined in \cite{nolan2017local}. We revisit that claim and its refutation here, and extend our understanding of McVittie spacetimes as black holes by showing that there are indeed timelike geodesics that extend to the event horizon in finite proper time. Along these geodesics (as is the case with ingoing radial null geodesics), all curvature and matter invariants remain finite in the limit as the horizon is reached. However, as noted in \cite{kaloper2010mcvittie}, there are differences between the case of a vanishing and a positive cosmological constant. We discuss these in detail below: briefly, there are singularities in second derivatives of the curvature tensor and in parallel propagated frame components of the curvature in the case of a vanishing cosmological constant that are not present when the cosmological constant is positive.

The remainder of this section is given over to recalling properties of McVittie spacetimes, focussing on the black hole interpretation. In the next section, we study timelike geodesics of McVittie spacetimes, and establish the result mentioned above, that one can indeed fall into a McVittie black hole. In the third section, we consider the limiting behaviour of various curvature quantities along causal geodesics that meet the horizon. We consider scalar curvature invariants of zero, first and second order in derivatives (so we consider in particular $R_{abcd}R^{abcd}, \nabla_e R_{abcd}\nabla^e R^{abcd}$ and $\nabla_f\nabla_e R_{abcd}\nabla^f\nabla^e R^{abcd}$). In Section 4, we calculate parallel propagated components of the curvature, and consider Jacobi fields - solutions of the (first order) geodesic deviation equation. For a positive cosmological constant, essentially everything is finite at the horizon. But for a vanishing cosmological constant, we find that the second order curvature scalar diverges (an effect first noted in \cite{kaloper2010mcvittie} for radial null geodesics), as do certain p.p.\ frame components of the curvature. In the concluding section, we argue that the results of Sections 3 and 4 cast a doubt on the usual interpretation of the fact that Jacobi fields are finite at the horizon - i.e.\ that extended test bodies remain intact on crossing the horizon. We use units with $c=G=1$ and follow the curvature conventions of \cite{wald1984general}. 

\subsection{McVittie spacetimes as black holes}

McVittie used co-moving coordinates to derive his solution of the Einstein equations \cite{mcvittie1933mass}. In these coordinates, and in the case of a spatially flat FLRW background, the line element reads
\be ds^2 = -\left(\frac{1-\frac{m}{2a\rho}}{1+\frac{m}{2a\rho}}\right)^2d\tau^2 + a^2\left(1+\frac{m}{2a\rho}\right)^4\left(d\rho^2+\rho^2d\Omega^2\right), \label{eq:lel-comov} \ee
where $a=a(\tau)$ is the scale factor of the FLRW background and $d\Omega^2=d\theta^2+\sin^2\theta d\phi^2$ is the line element of the unit 2-sphere $\mathbf{S}^2$. Global features of the spacetime are more conveniently analysed in proper-time/area radius coordinates, and the relevant coordinate transformation is given by 
\be t=\tau,\quad r = a\rho(1+\frac{m}{2a\rho})^2. \label{eq:coord-trans}\ee
In these coordinates, the line element can be written in the form \cite{nolan1999point}
\be ds^2 = -\alpha dt^2-2\beta dtdr + \gamma dr^2 + r^2 d\Omega^2 \label{eq:lel} \ee
where 
\begin{eqnarray}
\alpha &=& 1-\frac{2m}{r}-r^2H^2,\label{alpha-def} \\
\beta &=& r H\gamma^{1/2}, \label{beta-def} \\
\gamma &=& f^{-1},\quad f= 1-\frac{2m}{r}\label{gam-def}
\end{eqnarray}
Here $m$ is a constant parameter, assumed non-negative, and $H=H(t)$ is the Hubble function of the FLRW background. The metric functions satisfy the identity  
\begin{equation} \alpha\gamma+\beta^2 = 1,\label{al-be-ga-id}\end{equation} 
and the time coordinate $t$ is invariantly defined up to translation. 

The metric provides a solution of Einstein's equation with a perfect fluid source, with energy density and pressure given by 
\begin{equation} 8\pi\mu = 3H^2(t)-\Lambda,\qquad 8\pi P = -2H'(t)\gamma^{1/2}-3H^2(t)+\Lambda, \label{mcv-matter}\end{equation}
where $\Lambda$ is the cosmological constant. Here and throughout, a prime ($'$) represents the derivative with respect to argument. The fluid velocity vector field is given by 
\begin{equation} u = \gamma^{1/2}\pd{}{t}+\beta \gamma^{-1/2}\pd{}{r}.\label{u-mcv}\end{equation} The Newman-Penrose Weyl invariant is given by 
\be \Psi_2 = -\frac{m}{r^3}.\label{eq:psi2} \ee 
This is the only indepedent Weyl tensor component of the spacetime.  
The metric of (\ref{eq:lel})-(\ref{gam-def}), the matter quantities (\ref{mcv-matter}) and (\ref{u-mcv}) and the conformal curvature term (\ref{eq:psi2}) completely determine the curvature tensor of the spacetime.

The expansion of the fluid flow lines is given by $\theta=H$, and so the sign of $H$ distinguishes between a collapsing ($H<0$) and an expanding ($H>0$) spacetime (or region of spacetime). Note the occurence of a curvature singularity along $\{t=t_0<+\infty, r=2m\}$ except in the special case of vanishing $H'(t)$, and that the line element is defined only for $r>2m$. It follows that $t$ is a global time coordinate ($g^{ab}\nabla_at\nabla_bt = -(1-2m/r)^{-1}<0$), which we take to increase into the future. We recall the following features:

\begin{enumerate}
\item With $m=0$, equation (\ref{eq:lel}) gives the line element of a spatially flat FLRW universe with Hubble function $H(t)$.
\item With $H(t)=0$, the line element corresponds to  Schwarzschild spacetime with mass parameter $m$, and with $H(t)=H_0=$ constant, the line element corresponds to Schwarzschild-de Sitter spacetime with mass parameter $m$ and cosmological constant $\Lambda = 3H_0^2$.
\item The singularity at $r=2m$ is spacelike, and in the expanding case, forms a past boundary of the spacetime \cite{nolan1999point}.
\item In the expanding case and for $\Lambda\geq 0$, the spacetime contains a black hole horizon \cite{kaloper2010mcvittie,lake2011more,nolan2017local}.
\end{enumerate}

The black hole nature of the spacetime was first recognized in \cite{kaloper2010mcvittie}. In the case $\Lambda>0$, the authors argue that the spacetime can be extended across the black hole horizon to a portion of the maximally extended Schwarzschild de Sitter spacetime. In the case $\Lambda=0$, they argue that this extension is not possible, and so characterize the (would-be) black hole horizon as a null weak singularity. Further detail was added in \cite{lake2011more} who showed that for a particular (representative) case, the ($\Lambda>0$) extension incorporates both black and white hole horizons and the bifurcation 2-sphere. By integrating the null geodesic equations numerically, the black hole interpretation was argued to hold in the case $\Lambda=0$ for a dust-filled FLRW background in \cite{lake2011more}. The black hole interpretation was considered more generally in the case $\Lambda=0$ in \cite{nolan2017local} (see Section 10 of that paper, which also generalised a number of results to non-flat FLRW backgrounds). The black hole interpretation arises as follows - some further preamble is necessary to allow for precise statements. 

We assume an expanding FLRW background universe in which the weak energy condition holds, so that 
\be H(t)>0 \quad \hbox{ and }\quad H'(t)<0 \quad \hbox{ for all } \quad t>0.\label{eq:H-cons}\ee
We assume the asymptotic behaviour 
\be \lim_{t\to+\infty}(H(t),H'(t)) = (H_0,0)
\label{eq:H-lim} \ee
where
\be H_0= \left(\frac{\Lambda}{3}\right)^{1/2}, \quad  \Lambda\geq 0. \label{eq:H0-def}\ee
Notice that $H_0=0$ for $\Lambda=0$. Then \be \lim_{t\to+\infty} a(t)= +\infty \label{eq:scale-fac-lim} \ee in the case $\Lambda>0$ where $a$ is the scale factor of the FLRW background (so that $H(t)=a'(t)/a(t)$), and we assume that (\ref{eq:scale-fac-lim}) also holds in the case $\Lambda=0$. We also assume a Big Bang of the background at $t=0$: 
\be \lim_{t\to 0^+} \int_t^{t_0} H(u)du = +\infty \quad \hbox{ for all } t_0>0. \label{eq:bb} \ee

We note that these assumptions allow for a broad range of FLRW cosmological models (albeit restricted to the spatially flat case). This includes the standard $\Lambda$CDM model, for which the Friedmann equation may be written 
\be H^2 = \mathcal{H}_0^2\left(\Omega_r\left(\frac{a_0}{a}\right)^4+\Omega_m\left(\frac{a_0}{a}\right)^3+\Omega_\Lambda\right). \label{eq:fried} \ee
Here $\Omega_m, \Omega_r$ and $\Omega_\Lambda$ represent the proportions of the density in radiation, matter (incorporating baryonic and cold dark matter) and dark energy (modelled by a cosmological constant). $a_0$ and $\mathcal{H}_0$ are the values of the scale factor and Hubble function today, and recall that the curvature contribution $\Omega_K$ vanishes. See, for example, \cite{ellis2012relativistic}. We can calculate 
\be H' = -\frac12
\mathcal{H}_0^2\left(4\Omega_r\left(\frac{a_0}{a}\right)^4+3\Omega_m\left(\frac{a_0}{a}\right)^3\right).
\label{eq:fried-dh} \ee
Then for an expanding universe ($H>0$), we see that the assumptions (\ref{eq:H-cons})-(\ref{eq:bb}) all hold for this model, and allow for the presence ($H_0=\mathcal{H}_0\Omega_\Lambda^{1/2}>0$) or absence ($H_0=\mathcal{H}_0\Omega_\Lambda^{1/2}=0$) of dark energy. These assumptions also hold for FLRW spacetimes with $P_0=\omega\mu_0$ and $\Lambda=0$ for $-1<\omega\leq1$, including those models with accelerated expansion ($-1<\omega<-1/3$).

With these assumptions, and noting item (iii) above, we take the spacetime manifold to be $M=\mathcal{M}_2\times \mathbf{S}^2$ where 
\be \mathcal{M}_2=\{(t,r):t>0,r>2m\}. \label{eq:M2} \ee 
Noting that 
\be \alpha =g^{ab}\nabla_ar\nabla_br, \label{eq:alpha-grad} \ee we have a decomposition of the spacetime, in the expanding case, into a \textit{regular region}, an \textit{anti-trapped region} and a \textit{horizon} given by, respectively, 
\begin{eqnarray}
    \Omega_R&=&\{(t,r):\alpha(t,r)>0\},\label{eq:reg} \\
    \Omega_A&=&\{(t,r):\alpha(t,r)<0\},\label{eq:anti} \\
 \hor &=& \{(t,r):\alpha(t,r)=0\}. \label{eq:hor} 
 \end{eqnarray}
Let ${l}^\pm$ be tangent to the outgoing $(+)$ and ingoing $(-)$ radial null geodesics of the spacetime, and let $\theta^\pm$ be the corresponding null expansions. Then the regular region of the spacetime is characterised by $\theta^+>0, \theta^-<0$, the anti-trapped region has $\theta^+>0, \theta^->0$ and on the horizon $\theta^+>0,\theta^-=0$. (We could substitute ${l}^\pm(r)$ for $\theta^\pm$ in this last sentence.) The existence of a regular region of the spacetime is essential to the black hole interpretation, and in the case $\Lambda>0$, this implies the constraint 
\be mH_0 < \frac{1}{3\sqrt{3}}, \label{eq:mH0} \ee
which we assume henceforth. Then we define $r_-$ to be the least positive solution for $r$ of 
\be 1 - \frac{2m}{r}-r^2H_0^2 = 0. \label{eq:rminus-eqn} \ee
This gives $r_-=2m$ in the case of a vanishing cosmological constant, and $r_-\in(2m,3m)$ for $\Lambda>0$. Of course $r_-$ is the radius of the black hole horizon of Schwarzschild-de Sitter spacetime with $\Lambda=3H_0^2>0$ (and of Schwarzschild spacetime in the case $\Lambda=0$).  

The following result, established across \cite{kaloper2010mcvittie,lake2011more} and \cite{nolan2014particle}, shows that there is a boundary of the causal past of future null infinity that is accessible in finite affine time by certain null geodesics. This boundary is formed by the endpoints at $r=r_-$ of the geodesics of part (b) below, and these endpoints do not lie in the causal past of the set of future endpoints (future null infinity) of the geodesics of part (a). That is, the spacetime admits a black hole event horizon. See Propositions 4 and 5 of \cite{nolan2014particle} for complete and self-contained proofs. Following \cite{nolan2014particle}, we refer to a McVittie spacetime satisfying (\ref{eq:lel})-(\ref{gam-def}) and (\ref{eq:H-cons})-(\ref{eq:bb}) as an expanding McVittie spacetime with a Big Bang background. Figure 1 below shows the Penrose diagram in the case $\Lambda>0$.	

\begin{proposition}
Let $(M,g)$ be an expanding McVittie spacetime with a Big Bang background. 	
\begin{itemize}
\item[(a)] Every outgoing radial null geodesic of $(M,g)$ is future-complete, and 
\be \lim_{s\to+\infty} (t(s),r(s)) = (+\infty,+\infty) \label{eq:ORNGlim} \ee
where $s$ is an affine parameter along the geodesic. 
\item[(b)] For every ingoing radial null geodesic of $(M,g)$ with its initial point in $\Omega_R\cup\hor$, there exists a finite value $s_0$ of the affine parameter along the geodesic such that 
\be \lim_{s\to s_0^-} (t(s),r(s)) = (+\infty, r_-). \label{eq:IRNGlim} \ee
\hfill$\blacksquare$
\end{itemize} 
\end{proposition}

We conclude this section by noting that in the case $H_0=0$, we have $r_-=2m$. Thus (\ref{gam-def}) and (\ref{mcv-matter}) show the possibility that the pressure diverges in the limit as the black hole horizon is approached. It was shown in \cite{kaloper2010mcvittie} that this limit is in fact zero as the horizon is approached along ingoing radial null geodesics: this result was generalised in \cite{nolan2017local}. Thus the potentially divergent term $H'(t)\gamma^{1/2}$ vanishes in the limit as the horizon is approached along causal geodesics. However, looking closely, \cite{kaloper2010mcvittie} showed that the second order curvature term $\nabla_f\nabla_e R_{abcd}\nabla^f\nabla^e R^{abcd}$ \textit{diverges} in this limit - leading to their characterization of the putative black hole horizon  as a weak (or soft) null singularity in this case. The authors also argue that the extension that is constructed in the case $\Lambda>0$ is not possible in the case $\Lambda=0$. We discuss the limiting behaviour of various curvature terms in Section 3 below. 

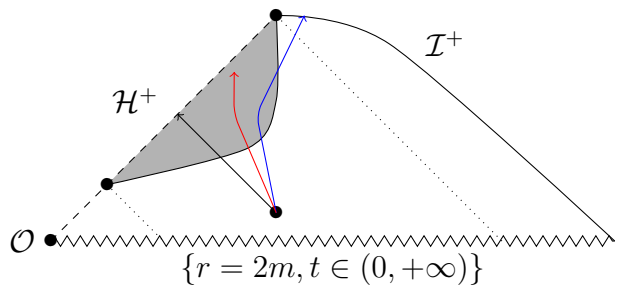
\begin{figure}
\begin{center}
\begin{tikzpicture}[scale=1.5]

\fill[gray!60] plot [smooth] coordinates{(0.5,0.5) (1.75,0.825) (2,1.25) (2,2)};

  \draw[dashed] (0,0) -- (2,2);
  \draw plot [smooth] coordinates {(2,2) (3,1.75) (5,0)};
    
  \draw[decorate,decoration={zigzag,segment length=5pt,amplitude=2pt}] (0,0) -- (5,0);
  
\draw plot [smooth] coordinates{(0.5,0.5) (1.75,0.825) (2,1.25) (2,2)};

\draw[dotted] (0.5,0.5) -- (1,0);
\draw[dotted] (4,0) -- (2,2);

  \node at (-0.25,0) {$\mathcal{O}$};
  \node at (0.75,1.25) {$\mathcal{H}^+$};
\node at (3.5,1.75) {$\mathcal{I}^{+}$};
 \node at (0,0){$\bullet$};
 \node at (0.5,0.5){$\bullet$};
 \node at (2,0.25){$\bullet$};
 \node at (2,2){$\bullet$};
 \node at (2.5,-0.25){$\{r=2m,t\in(0,+\infty)\}$};

\draw[->] (2,0.25) -- (1.125,1.125);
\draw[rounded corners,->,blue] (2,0.25) -- (1.825,1.125) -- (2.25,2);
\draw[rounded corners,->,red] (2,0.25) -- (1.625,1.125) -- (1.6275,1.5);

\end{tikzpicture}
\caption{Penrose diagram of a McVittie spacetime with $\Lambda>0$. $\mathcal{I}^+$, future null infinity, is spacelike everywhere in this case. The regular region is shaded: the unshaded region is anti-trapped. $\mathcal{H}^+$ is the black hole horizon - the future boundary of the causal past of $\mathcal{I}^+$. $t$ increases from left to right along the past singularity at $r=2m$. The left-most arrow (black) represents an ingoing radial null geodesic, along which $(t,r)\to(+\infty,r_-)$ as $\mathcal{H}^+$ is approached. The right-most arrow (blue) represents a fluid flow line: $(t,r)\to(+\infty,+\infty)$ as proper time increases without bound along these worldlines. The key question we address in this paper is whether or not timelike geodesics (represented by the central arrow (red)) can reach $\mathcal{H}^+$ in finite proper time. See \cite{kaloper2010mcvittie},\cite{lake2011more} and \cite{nolan2014particle} for further details.} 
\end{center}
\end{figure}

\section{Falling into a McVittie black hole: timelike geodesics}

We begin the discussion of particle trajectories by considering the fluid particles of the spacetime. The unit timelike tangent for these particles is given by (\ref{u-mcv}). These flow lines are not geodesic, but nevertheless the question of the asymptotic behaviour as $t\to+\infty$ of these particles is of interest. It was first noted in \cite{lake2011more} that $r\to+\infty$ as $t\to+\infty$ along the flow lines in a particular (but representative) case with $\Lambda>0$. We see here that this generalises to the whole class considered in this paper. To see this, we return to the comoving coordinates $(\tau,\rho)$ originally used by McVittie \cite{mcvittie1933mass}. 
Recall from (\ref{eq:coord-trans}) that 
\be r=a\rho\left(1+\frac{m}{2a\rho}\right)^2.\label{eq:area-rho} \ee
Since the comoving coordinate $\rho$ is constant (and positive) along each fluid flow line, we immediately see from (\ref{eq:scale-fac-lim}) that $r\to+\infty$ as $t\to+\infty$ along the flow lines. 

This simple result explains a curious feature of McVittie spacetimes: ingoing radial null geodesics detect no matter at the black hole event horizon. This is immediate in the case of the matter density $\mu$ - which is spatially homogeneous in McVittie spacetime. As we see from (\ref{mcv-matter}) and (\ref{eq:H-lim}), in the limit as we approach the horizon along an ingoing radial null geodesic, we have 
\be \lim_{t\to+\infty} \mu(t) = 0. \label{eq:mu-lim} \ee
The same result also holds for the pressure. This is immediate from (\ref{mcv-matter}) and (\ref{eq:H-lim}) in the case $\Lambda>0$ (the horizon is at $r=r_->2m$), but is a little more subtle in the case $\Lambda=0$. Here, the fact that the ingoing radial null geodesic approaches the horizon from the regular region $\Omega_R$ of the spacetime is essential to this conclusion (see \cite{kaloper2010mcvittie}, and Section 10 of \cite{nolan2017local} for a rigorous argument). The observation above about the fluid flow lines provides an explanation of this: we can think of the ingoing null geodesics as crossing \textit{all} of the fluid flow lines before reaching the horizon. 

These points provide a clear refutation of the claim in \cite{Poplawski_2025} regarding McVittie spacetime that ``the curvature scalar and pressure at its event horizon are infinite". The claim appears to be based on the observation that at finite $t$, the pressure diverges at $r=2m$, which is referred to in \cite{Poplawski_2025} as being the location of the (putative) event horizon of the spacetime. As we see from above, (i) this is not the location of the event horizon in the case $\Lambda>0$ and (ii) while this is the location of the event horizon in the case $\Lambda=0$, there is no divergence at the level of the curvature when we approach the horizon along ingoing radial null geodesics. It is immediate that there is no divergence in the case $\Lambda>0$, as the pressure remains finite for all $t>0$ at $r=r_->2m$. Of course this is not the main point of \cite{Poplawski_2025}, which provides an alternative model of a black hole in an expanding universe. 

These considerations give rise to the other questions raised in the introduction. Can other particles fall into a McVittie black hole? If so, how do the curvature and matter invariants behave along the particle world lines as the horizon is reached? We answer the first question in the affirmative: in every McVittie spacetime, there are timelike geodesics that extend to the black hole event horizon in finite proper time. We find that the values along these geodesics of curvature and matter terms are finite in the limit as the horizon is reached, answering the second question. However, other curvature terms do indeed diverge, as detailed in Sections 3 and 4 below.  

The equations governing timelike geodesics of McVittie spacetime are 
\begin{eqnarray} 
\ddot{t} &=& -\left(1-\frac{3m}{r}\right)H\gamma^{1/2}\dot{t}^2-\frac{2m}{r^2}\gamma\dot{t}\dot{r}+\gamma^{1/2}H,\label{eq:geo-t2} \\
\ddot{r} &=& r\gamma^{-1/2}H'\dot{t}^2 -\left(\frac{m}{r^2}-rH^2\right)+\left(1-\frac{3m}{r}\right)\frac{\ell^2}{r^3},\label{eq:geo-r2} 
\end{eqnarray}
and we have the first integral 
\be -\alpha\dot{t}^2-2rH\gamma^{1/2}\dot{t}\dot{r}+\gamma\dot{r}^2 +\frac{\ell^2}{r^2}= -1. \label{eq:geo-fi} \ee
 The overdot represents the derivative with respect to proper time and $\ell$ is the conserved angular momentum per unit mass along the geodesic. We note that to obtain the forms (\ref{eq:geo-t2}) and (\ref{eq:geo-r2}), we use the Euler-Lagrange equations and then eliminate the terms in $\dot{r}^2$ using (\ref{eq:geo-fi}). Since $t$ is a global time coordinate on the spacetime, $\dot{t}\neq0$ and we consider future-directed geodesics so that $\dot{t}>0$. 

The following inequality that holds in the regular region $\Omega_R$ of the spacetime is trivial to prove, but plays an important role below and so is highlighted:

\begin{lemma} If $\alpha>0$, then 
\be rH^2-\frac{m}{r^2} < \frac{1}{r}\left(1-\frac{3m}{r}\right). \label{eq:3m-ineq}\ee
\hfill$\blacksquare$
\end{lemma}

Initially ingoing ($\dot{r}<0$) timelike geodesics that originate at an event in $\Omega_R$ sufficiently close to the horizon remain ingoing, remain in $\Omega_R$ and remain close to the horizon: 

\begin{lemma} If 
\be \dot{r}<0,\quad \alpha>0,\quad r<3m \label{eq:stay} \ee
at an initial point on a timelike geodesic, then these inequalities continue to hold along the geodesic on the right-maximal interval of existence. 
\end{lemma} 

\noindent\textbf{Proof:} Suppose the geodesic meets a point $p$ at which $\dot{r}=0$. 
We have $H'<0$ (\textit{cf.} (\ref{eq:H-cons})), and if $\alpha>0$ and $r<3m$, (\ref{eq:geo-r2}) and (\ref{eq:3m-ineq}) show that $\left.\ddot{r}\right|_p<0$. So $r$ cannot have a minimum turning point,  $\dot{r}$ remains negative and $r$ remains below $3m$. On the other hand, suppose that the geodesic meets a point $p\in\hor=\{\alpha=0\}$. Using (\ref{al-be-ga-id}) and $H>0$, we have $\beta|_p=1$. Then (\ref{eq:geo-fi}) gives 
\be \left.\gamma\dot{r}^2-2\dot{t}\dot{r}+1+\frac{\ell^2}{r^2}\right|_p = 0, \label{eq:geo-al-zero} \ee
and so $\left.\dot{r}\right|_p>0$. Thus $r$ cannot reach a minimum before the geodesic leaves the regular region, and the geodesic cannot leave the regular region before it attains a minimum of $r$. So $r$ must continue to decrease along the geodesic (and consequently stays below $3m$), and the geodesic must remain in the region $\alpha>0$. \hfill$\blacksquare$

We restrict henceforth to timelike geodesics satisfying the conditions established in this lemma. (Our aim is to show that \textit{some} observers can fall into a McVittie black hole rather than give a comprehensive account of the timelike geodesics of the spacetime.)   

The convexity properties of these geodesics are also uniform. The following result follows immediately for the geodesics of Lemma 2.2 by inspection of the terms on the right hand sides of (\ref{eq:geo-t2}) and (\ref{eq:geo-r2}). 
\begin{corollary} If $\dot{r}<0$, $\alpha>0$ and $r<3m$ at an initial point of a timelike geodesic, then 
\be \ddot{t}>0,\quad \ddot{r}<0 \label{eq:convexity} \ee
on the right-maximal interval of existence. \hfill$\blacksquare$ 
\end{corollary} 

Our main result is the following.  

\begin{theorem}
    Consider a future-directed timelike geodesic of an expanding McVittie spacetime with a Big Bang background satisfying the initial conditions at proper time $s=0$
    \be \left.r\right|_{s=0}<3m,\quad \left.\dot{r}\right|_{s=0}<0,\quad \left.\alpha\right|_{s=0}>0. \label{eq:ics} \ee
    Then there exists $s_0\in(0,\infty)$ such that 
    \be \lim_{s\to s_0^-}r(s) = r_-,\quad \lim_{s\to s_0^-} t(s) = +\infty. \label{eq:limits} 
    \ee
\end{theorem}

We need to recall the following technical lemma (\textit{cf.} Lemma 5 of \cite{nolan2014particle}): 

\begin{lemma}
    There exists a $C^1$ function $w:(H_0,+\infty)\to (-\infty,0)$ such that $H'(t) = w(H(t))$ for all $t\in(0,+\infty)$, and $\lim_{H\to H_0^+}w(H)=0$. \hfill$\blacksquare$
\end{lemma}

For convenience, we also state a standard result of dynamical systems which is central to the argument below. See e.g.\  Theorem 2 of Section 2.4 of \cite{perkodifferential}:

\begin{theorem} Let $F\in C^1(E,\real^n)$ where $E$ is an open subset of $\real^n$, and let $(t_0,t_\omega)$ be the right-maximal interval of existence of the initial value problem
\be \frac{du}{dt} = F(u),\quad u(t_0)=u_0\in E.\label{eq:Thm-ivp} \ee
If $t_\omega<+\infty$ and $K$ is a compact subset of $E$, then the solution of (\ref{eq:Thm-ivp}) exits $K$. That is, there exists $t\in(t_0,t_\omega)$ such that $u(t)\not\in K$. \hfill$\blacksquare$    
\end{theorem}

\noindent\textbf{Proof of Theorem 2.1:} Define $u=\dot{t}>0$. By Lemma 2.2, the inequalities of (\ref{eq:ics}) remain true along the geodesic while $u$ remains finite. Then using $t$ as the parameter along the geodesic, choosing the appropriate root of (\ref{eq:geo-fi}) and applying Lemma 2.3, we can write the relevant geodesic equations in the form
\begin{eqnarray} 
\frac{dr}{dt} &=& \frac{1}{\gamma}\left(\beta-\left(1-\frac{\gamma}{u^2}\left(1+\frac{\ell^2}{r^2}\right)\right)^{1/2}\right),\label{eq:rdot} \\
\frac{du}{dt} &=& -\left(1-\frac{3m}{r}\right)H\gamma^{1/2}u-\frac{2m}{r^2}  \left(\beta-\left(1-\frac{\gamma}{u^2}\left(\delta+\frac{\ell^2}{r^2}\right)\right)^{1/2}\right)u \nonumber \\
&& + \gamma^{1/2}\frac{H}{u},\label{eq:udot} \\
\frac{dH}{dt} &=& w(H). \label{eq:hdot}
\end{eqnarray}
Here, $\delta=1$ for timelike geodesics and $\delta=0$ for null geodesics. (We don't consider the latter in the present theorem: $\delta$ is included for later convenience.) We use $du/dt = \dot{u}/\dot{t}=\ddot{t}/\dot{t}$, and substitute for $\dot{r}/\dot{t} = dr/dt$ by using the first equation in the second. We note that since $\dot{r}<0$, we can use (\ref{eq:geo-fi}) to show that 
\be 1-\frac{\gamma}{u^2}\left(1+\frac{\ell^2}{r^2}\right) >\beta^2 >0. \label{eq:positive-coefficient} \ee This casts the timelike geodesic equations as an autonomous system with independent variable $t$, unknown $(r,u,H)$ and $C^1$ right hand side. To prove the theorem, we show that solutions of this system exist globally in $t$, that the limits corresponding to (\ref{eq:limits}) hold as $t\to +\infty$, and then finally that $t$ diverges to $+\infty$ in finite proper time. 

Let $t_0=\left.t\right|_{s=0}$, and let $[t_0,t_\omega)$ be the right-maximal interval of existence of (\ref{eq:rdot})-(\ref{eq:hdot}) subject to 
\be r(t_0) = \left.r\right|_{s=0}<3m,\quad u(t_0) = \left.\dot{t}\right|_{s=0}>0\label{eq:ics-t}\ee
and with $\alpha(t_0,r(t_0))>0$. The initial condition for $H$ is the known value $H(t_0)>H_0$. Suppose that $t_\omega<+\infty$: we show that this leads to a contradiction.

We note first that 
\be H(t) \geq H(t_\omega) > H_0,\quad t_0<t\leq t_\omega. \label{eq:Hbounds} \ee This follows as $H$ is monotone decreasing, and reaches $H_0$ only in the limit as $t\to+\infty$. Since $\alpha$ remains positive on the geodesic, it follows that $r$ must remain strictly above $r_-$ for all $t<t_\omega$. Thus there exists $\epsilon>0$ such that 
\be (r(t),H(t))\in [r_-+\epsilon,r(t_0)] \times [H_0+\epsilon,H(t_0)]\quad \hbox{for all} \quad t\in (t_0,t_\omega).\label{eq:K1} \ee
Since $u$ is strictly increasing ($\ddot{t}>0,\dot{t}>0$), we know that $u$ cannot return to the value $u(t_0)$ on $(t_0,t_\omega)$. For $\lambda>u(t_0)$, define
\be K_\lambda = [r_-+\epsilon,r(t_0)] \times [u(t_0),\lambda]\times [H_0+\epsilon,H(t_0)].\label{eq:Kdef} \ee

Applying Theorem 2.2, we see that the trajectory $(r(t),u(t),H(t))$ must exit $K_\lambda$ (since $\epsilon>0$, we can find an open subset $E$ of $\real^3$ such that the right-hand sides of (\ref{eq:rdot})-(\ref{eq:hdot}) are $C^1$ on $E$, and such that $K_\lambda\subset E$). The comments above show that this can only happen at the boundary $u=\lambda$, and so there must exist $t\in(t_0,t_\omega)$ such that $u(t)>\lambda$. Since $u$ is monotone and $\lambda$ can be arbitrarily large, this implies that 
\be \lim_{t\to t_\omega^-} u(t) =+\infty. \label{eq:u-lim-infinty} \ee
But then (\ref{eq:udot}) gives 
\be \frac{du}{dt} \sim \left[ -\left(1-\frac{3m}{r}\right)H\gamma^{1/2}-\frac{2m}{r^2}  \left(\beta-1\right)\right]_{t=t_\omega}u,\quad t\to t_\omega^-.
\label{eq:dudt-asymp}
\ee
The coefficient here is a limiting value that must exist and be finite by monotonicity of $r$ and $H$. This asymptotic relation is inconsistent with divergence of $u$ as $t\to t_\omega$. Hence the assumption that $t_\omega<+\infty$ leads to a contradiction, and so the solution of (\ref{eq:rdot})-(\ref{eq:hdot}) with (\ref{eq:ics-t}) exists globally (to the future) in $t$. 

Next, we observer that all three terms on the right hand side of (\ref{eq:udot}) are positive, and so we have 
\be u\frac{du}{dt} > \gamma^{1/2}H >\gamma_0^{1/2} H,\quad t>t_0, \label{eq:u-udot-ineq} \ee
where we use $\gamma_0=\gamma(r(t_0))$ and the fact that $\gamma=(1-2m/r)^{-1}$ is monotone increasing. Integrating yields
\be u^2 >(u(t_0))^2 + 2\gamma_0^{1/2}\log\left(\frac{a(t)}{a(t_0)}\right),\quad t\to t_0, \label{eq:u2-bound} \ee and so 
\be \lim_{t\to+\infty} u(t) =+\infty. \label{eq:u-inf-lim} \ee

We prove that $\lim_{t\to+\infty}\alpha=0$ as follows. Appealing again to positivity of each term on the right hand side of (\ref{eq:udot}), and using (\ref{eq:rdot}), we have 
\be u^{-1}\frac{du}{dt} > -\frac{2m}{r^2}\left(1-\frac{2m}{r}\right)^{-1}\frac{dr}{dt}. \label{eq:u-udot-bound2} \ee
Integrating yields
\be u\left(1-\frac{2m}{r}\right)>u(t_0)\left(1-\frac{2m}{r(t_0)}\right),\quad t>t_0, \label{eq:u-gam-bound} \ee
and so 
\be \frac{\gamma}{u^2} < \frac{1}{u(t_0)\left(1-\frac{2m}{r(t_0)}\right)u}. \label{eq:u-gam-bound2} \ee
Hence 
\be \lim_{t\to+\infty} \frac{\gamma}{u^2} = 0. \label{eq:u-gam2-lim} \ee
(This is immediate in the case where $H_0>0$, but when $H_0=0$, we have $\gamma\to+\infty$ in the limit.) Since $r$ is monotone decreasing and bounded below on $(t_0,+\infty)$, we must have $\lim_{t\to+\infty} dr/dt=0$ and so we can use (\ref{eq:rdot}), (\ref{eq:u-gam2-lim}) and the identity (\ref{al-be-ga-id}) to obtain 
\begin{eqnarray} 
0 &=& \lim_{t\to+\infty} \frac{dr}{dt} \nonumber \\
&=& \lim_{t\to+\infty} \frac{\beta-1}{\gamma} \nonumber \\
&=& \lim_{t\to+\infty} \frac{\alpha}{\beta+1} \nonumber \\
&=& \lim_{t\to+\infty} \frac{\alpha(1-2m/r)^{1/2}}{rH+(1-2m/r)^{1/2}}. 
\label{eq:al-lim} 
\end{eqnarray}

For both cases of $H_0$ (positive and zero), this gives 
\be \lim_{t\to+\infty} \alpha = 0. \label{eq:alpha-limit} \ee
We know that $r$ is decreasing and bounded below by $r_-$. Hence the limit $r_\omega = \lim_{t\to+\infty} r(t)$ exists and satisfies $r_\omega\in[r_-,3m)$. Equation (\ref{eq:alpha-limit}) tells us that we must have 
\be 1-\frac{2m}{r_\omega}-r_\omega^2H_0^2 = 0, \label{eq:r-omega} \ee
and so 
\be \lim_{t\to+\infty} r(t) = r_-. \label{eq:r-inf-lim} \ee

The last step in the proof is to show that $t\to+\infty$ in a finite amount of proper time. By monotonicity, this is equivalent to showing that $r\to r_-$ in finite proper time. This follows by integrating $\ddot{r}<0$ twice (see (\ref{eq:convexity})), and using $\dot{r}<0$. \hfill$\blacksquare$

\section{Curvature at the horizon}

We have established the existence of timelike geodesics that reach the black hole horizon in finite proper time. In this section, we consider the behaviour of curvature and matter invariants along these geodesics - and along null geodesics - as they meet the black hole horizon. We consider zero-order invariants constructed from the Riemann tensor,  such as $R, R_{ab}R^{ab}, R_{abcd}R^{abcd}$ and $C_{abcd}C^{abcd}$ (where $C_{abcd}$ is the Weyl tensor), and we also consider the first- and second-order curvature invariants $\nabla_eR_{abcd}\nabla^eR^{abcd}$ and $\nabla_f\nabla_eR^{abcd}\nabla^f\nabla^eR^{abcd}$ previously considered in \cite{kaloper2010mcvittie} and \cite{lake2011more}. We deal with both cases $H_0>0$ and $H_0=0$. In the latter case, \cite{kaloper2010mcvittie} give a heuristic argument  that zero- and first-order invariants remain finite as the horizon is approached along radial null geodesics, but that the second order term diverges in this limit. The conclusion in relation to the second order invariant is disputed in \cite{lake2011more}. In this section, we consider the general form of these invariants, and identify a zero-order quantity whose limiting behaviour determines the behaviour of all zero-, first- and second-order invariants. This applies to the limiting behaviour along any causal geodesic that meets the black hole horizon in finite affine (or proper) time. Our conclusions establish rigorously the claim of \cite{kaloper2010mcvittie}, that there is an extremely mild singularity along the horizon in the case $H_0=0$: zero- and first-order invariants are finite in the limit, but the second-order term diverges. 

\subsection{Curvature invariants.}

For the calculations of this section, it is convenient to use the Ricci decomposition of the Riemann tensor (see e.g.\ \cite{wald1984general}):
\be R_{abcd} = C_{abcd} + g_{a[c}R_{d]b}-g_{b[c}R_{d]a} -\frac{R}{3}g_{a[c}g_{d]b}. \label{eq:riem-decomp} \ee
Since the Weyl tensor is trace-free on all index pairs, this leads to a convenient form for the Kretschmann scalar: 
\be R_{abcd}R^{abcd} = C_{abcd}C^{abcd} + 2R_{ab}R^{ab} - \frac13R^2. \label{eq:kretsch0} \ee
Since the metric commutes with the metric connection $\nabla_a$, it follows immediately that the same decomposition applies to first- and second-order derivatives of the Riemann tensor: 
\be \nabla_eR_{abcd}\nabla^eR^{abcd} = \nabla_eC_{abcd}\nabla^eC^{abcd} + 2\nabla_e R_{ab}\nabla^eR^{ab} - \frac13\nabla_e R\nabla^e R, \label{eq:kretsch1} 
\ee
\begin{eqnarray} 
 \nabla_f\nabla_eR_{abcd}\nabla^f\nabla^eR^{abcd} &=& D2\textrm{Weyl} + D2\textrm{Ric},
 \label{eq:kretsch2} \end{eqnarray}

 where
 \begin{eqnarray}
     D2\textrm{Weyl}&=&
 \nabla_f\nabla_eC_{abcd}\nabla^f\nabla^eC^{abcd},\label{eq:d2weyl-def} \\
 D2\textrm{Ric} &=& 2\nabla_f\nabla_e R_{ab}\nabla^f\nabla^e R^{ab} - \frac{1}{3}\nabla_f\nabla_e R\nabla^f\nabla^eR. \label{eq:d2ric-def} \end{eqnarray}


In this perfect fluid, spherically symmetric (and hence Petrov Type D) spacetime, the zero-order curavture invariants are all sums of products and powers of the Weyl invariant $\Psi_2$, the matter density $\mu$, the pressure $P$ and the cosmological constant $\Lambda$, and for convenience we recall 
\be \Psi_2 = -\frac{m}{r^3},\quad  8\pi\mu = 3H^2-\Lambda,\quad 8\pi P = -2H'\gamma^{1/2}-3H^2+\Lambda.
\label{eq:zero-terms}\ee
We calculate 
\be R_{abcd}R^{abcd} = 48\frac{m^2}{r^6}+24H^4-24H^2H'\gamma^{1/2}+12(H'\gamma^{1/2})^2,\label{eq:McV-K0} \ee 
Evidently, there is a zero-order scalar curvature divergence if and only if 
\be J  := H'\gamma^{1/2} = H'\left(1-\frac{2m}{r}\right)^{-1/2}, \label{eq:J-def} \ee
diverges in the limit as the horizon is approached. 

The calculation of the first order term (\ref{eq:kretsch1}) is simplified by the fact that the McVittie fluid flow is shear-free (and twist-free). We find 
\be \nabla_au_b =-u_aa_b+H(g_{ab}+u_au_b),\label{eq:du} \ee
where the acceleration is found to be 
\be a = \frac{m}{r^2}\pd{}{r}.\label{eq:acc-def} \ee
This has norm (squared) 
\be  a\cdot a = g_{ab}a^aa^b= \frac{m^2}{r^4}\gamma.\label{eq:ada} \ee


We find 
\begin{eqnarray} \nabla_eR_{abcd}\nabla^eR^{abcd} &=& 720\frac{m^2}{r^8}\alpha+28\frac{m^2}{r^4}\gamma^2(H')^2-12\gamma^2(H'')^2\nonumber \\ && -12(12-52\frac{m}{r}+57\frac{m^2}{r^2})\gamma^3H^2(H')^2 \nonumber \\ && -24(2-5\frac{m}{r})\gamma^{5/2}HH'H''.\label{eq:K1-McV} \end{eqnarray}

The (lengthy) expressions for the second-order terms $D2\textrm{Weyl}$ and $D2\textrm{Ric}$ are given in Appendix A. These expressions take the general form
\begin{eqnarray} D2\textrm{Weyl} &=& \sum_{i_k} P_{i_1}\gamma^{i_2/2}H^{i_3}(H^{(i_4)})^{i_5},\label{D2Weyl-form} \\
D2\textrm{Ric} &=& \sum_{j_k} P_{j_1}\gamma^{j_2/2}H^{j_3}(H^{(j_4)})^{j_5} \label{D2Ric-form} 
\end{eqnarray}
where $i_k, j_k, k=1,..,5$ are non-negative integers and $P_{i_1}$ and $P_{j_1}$ are polynomials in $m/r$ of degree $i_1, j_1$ respectively which are non-vanishing at $r=2m$.

\subsection{The case $H_0>0$.}

As we have seen, $t\to+\infty$ and $r\to r_-$ as the horizon is approached. Then we can immediately deal with the case of a positive cosmological constant. In this case, the horizon radius is $r_->2m$ and $\gamma$ has a finite limit. Hence we see that all of the curvature scalars considered above have finite limits in the approach to the black hole horizon. In the case of the second order invariant (\ref{eq:kretsch2}), this is subject to the additional assumption that $H\in C^3(0,+\infty)$ with $\lim_{t\to+\infty} H^{(3)}(t)=0$.


\subsection{The case $H_0=0$.}

The case $H_0=0$ is a bit more delicate. We carry out the following steps to calculate the limits of the various curvature invariants. 

As we will see, knowing that $\alpha$ remains positive along a causal geodesic meeting the black hole horizon (and vanishes in the limit) is central to the argument. We know this to be the case for the timelike geodesics of Theorem 2.1. In fact this theorem describes all such observers: any causal geodesic that reaches the black hole horizon in finite proper time must satisfy the hypotheses of Theorem 2.1. This is established in Lemma 3.1 below. 

Next, we must consider the behaviour of the Hubble function in the limit $t\to+\infty$. As in the case $H_0>0$, we see from (\ref{eq:ddweyl-squared}) and (\ref{eq:D2Ric-squared}) that third derivatives are needed for the calculation of the second order curvature invariant (\ref{eq:kretsch2}). Making a mild assumption on the equation of state in the FLRW background allows us to determine the asymptotic behaviour of the first three derivatives of $H$: these are required for the calculation of the limits. See Lemma 3.2 below. 

We then see that the limiting behaviour of the curvature invariants is completely determined by the limiting behaviour of a single quantity: $K=H\gamma^{1/2}$. We prove that this has the finite limit $1/2m$ as the horizon is approached along any causal geodesic (Lemma 3.3). These steps allow us to prove Proposition 3.1, which establishes that the zero- and first-order curvature terms are finite in the limit, but that the second order term (\ref{eq:kretsch2}) diverges along all causal geodesics as the horizon is reached. 

\subsubsection{Universal behaviour of causal geodesics approaching the horizon.}

\begin{lemma}
    Let $C$ be a causal geodesic of an expanding McVittie spacetime with a Big Bang background with $H_0=0$. Suppose that $C$ reaches the black hole horizon in finite proper time (affine time for null geodesics). Then there exists an event on the geodesic with \be r<3m, \quad \dot{r}<0,\quad \hbox{and} \quad \alpha>0.
    \label{eq:all-geos} \ee
\end{lemma}

\noindent\textbf{Proof:} The black hole horizon is located at $r=2m$, the minimum value of $r$ on the spacetime. Thus $r$ must have (at the least) a sequence of decreasing values as proper time $s$ increases along the geodesic. If $\alpha\leq0$ at any such point, we see from (\ref{eq:geo-fi}) that 
\be 2\beta\dot{t}\dot{r}=-\alpha\dot{t}^2+\gamma\dot{r}^2+\delta+\frac{\ell^2}{r^2} >0,\label{eq:al-neg} \ee
and so $\dot{r}>0$. As in (\ref{eq:rdot}), $\delta=+1$ for timelike geodesics, and $0$ for null geodesics. So to reach the horizon, the geodesic must lie in the region $\alpha>0$ (for values of proper time sufficiently close to the time at which the geodesic does so). We clearly cannot have $r\geq 3m$ on this segment of the geodesic, and the conclusion follows: there must be points on the geodesic with $\dot{r}<0,\alpha>0$ and $r<3m$. \hfill$\blacksquare$

Now let $C$ be a causal geodesic that meets the black hole horizon at proper/affine time $s_0<+\infty$. By Proposition 3.2 and Lemma 2.2, there exists $s_*<s_0$ such that the inequalities (\ref{eq:all-geos}) hold for all $s\in(s_*,s_0)$ on $C$.  Likewise, Proposition 5 of \cite{nolan2014particle} and the trivial extension of Proposition 3.2 to the null case shows that these inequalities also hold along any null geodesic that meets the black hole horizon in finite affine time $s=s_0<+\infty$. Recall also that we can take the time coordinate $t$ to be a parameter along the geodesic. So for $C$ as in Lemma 3.1, we have the following:

\begin{lemma} There exists $t_0<+\infty$ such that 
\be \alpha(t):=\left.\alpha\right|_C>0\quad \hbox{for all } t>t_0.\label{eq:alpha-pos} \ee
\end{lemma}

\subsubsection{The background equation of state and asymptotics of $H^{(n)}$.}


It is clear from the forms of the curvature invariants above that their limiting behaviour at the horizon depends on the relative scaling of $\gamma$ with $H$ and its derivatives. We know already that $H$ and $H'$ vanish in the limit: the key question is whether or not this suppresses the divergence of $\gamma$. So more precise asymptotics are required. We can do this by considering the equation of state of the background FLRW spacetime. 

Taking $m=0$ in (\ref{mcv-matter}), the density $\mu_0$ and pressure $P_0$ of the background are given by 
\be 8\pi\mu_0 = 3H^2-\Lambda,\quad 8\pi P_0 = -2H'-3H^2+\Lambda. \label{eq:FLRW}\ee
We assume that there is a barotropic equation of state of the form 
\be P_0=g(\mu_0),\quad g:\real_+\to\real\label{eq:eos} \ee where $g$ is differentiable at the origin (i.e. in the limit of zero density). Define the parameter 
\be \kappa = \frac32(1+g'(0))>0, \label{eq:kappa-def}\ee 
which relates to the sound speed at zero density. The sign follows from the weak energy condition. It is straightforward to establish the following generalisation of Lemma 6 of \cite{nolan2014particle} (use $g'(0)=\lim_{t\to+\infty} P_0/\mu_0$ for the first line, and then apply l'Hopital's rule twice):

\begin{lemma} Let $H\in C^3((0,+\infty),\real_+)$ be the Hubble function of an FLRW background satisfying (\ref{eq:H-cons})-(\ref{eq:scale-fac-lim}) and the conditions (\ref{eq:eos}) and (\ref{eq:kappa-def}) on the equation of state. Then  
\begin{eqnarray}
    H' &\sim& -\kappa H^2,\label{eq:Hp-lim} \\
    H'' &\sim& 2\kappa^2H^3,\label{eq:Hpp-lim} \\
    H^{(3)} &\sim& -4\kappa^3H^4,\label{eq:Hppp-lim} 
\end{eqnarray} 
where the asymptotic behaviour holds in the limit $t\to+\infty$ in each case. \hfill$\blacksquare$
\end{lemma} 

\begin{comments} We note that the condition (\ref{eq:kappa-def}) does not hold for the $\Lambda$CDM Hubble function of (\ref{eq:fried}). However, in that case we can calculate directly that the asymptotic relations (\ref{eq:Hp-lim})-(\ref{eq:Hppp-lim}) continue to hold.\end{comments}

\subsubsection{Some key limits.}
Recall that $J=H'\gamma^{1/2}$ and define
\be K=H\gamma^{1/2}. \label{eq:k-def} \ee
By applying Lemma 3.2 and (\ref{eq:Hp-lim}) of Lemma 3.3, we can write down the following. 

\begin{lemma} 
Let $C$ be a causal geodesic that meets the black hole horizon in finite affine (or proper) time. Then 
\be \lim_{t\to+\infty} \left. J\right|_C = 0. \label{eq:J-lim-zero} \ee
and there exists $t_0$ such that 
\be \left.K\right|_C \in (0,\frac{1}{2m})\quad \hbox{for all }\quad t>t_0.  \label{eq:k-bound} \ee
\hfill$\blacksquare$
\end{lemma}


We can say more about the crucial quantity $K$:

\begin{lemma} 
Let $C$ be as in Lemma 3.4. Then 
\be \lim_{t\to+\infty} \left.K\right|_C = \frac{1}{2m}.\label{eq:K-lim}\ee
\end{lemma}

\noindent\textbf{Proof:} As seen, we can use the global time coordinate as a parameter along the geodesic $C$. A direct calculation using (\ref{eq:udot}) yields
\begin{eqnarray}
    \frac{dK}{dt} &=& \frac{H'}{H}K-\frac{m}{r}K^2+\frac{m}{r^2}K\left(1-\frac{\gamma}{u^2}\left(\delta+\frac{\ell^2}{r^2}\right)\right)^{1/2} . \label{eq:dK} 
\end{eqnarray}
We extract the leading order behaviour of each term on the right hand side which allows us to write
\be \frac{dK}{dt} = \frac{K}{4m} - \frac12K^2 + \lambda\frac{K}{4m} \label{eq:dk-leading} \ee
where
\be \lambda = 4m\frac{H'}{H}+\left(2m-\frac{4m^2}{r}\right)K+\frac{4m^2}{r^2}\left(1-\frac{\gamma}{u^2}\left(\delta+\frac{\ell^2}{r^2}\right)\right)^{1/2}-1. \label{eq:lam-def} \ee
Using (\ref{eq:Hp-lim}), (\ref{eq:u-gam2-lim}), the fact that $r\to2m$ at the horizon and the fact that $K$ is bounded in the limit as the horizon is approached, we see that $\lambda:[t_0,+\infty)$  is continuous with 
\be \lim_{t\to\infty} \lambda(t) = 0. \label{eq:lam-lim} \ee 
Define $v$ by 
\be K=\frac{v}{2m},\quad v \in (0,1),\quad t>t_0\label{eq:v-def} \ee
where $t_0$ is as in Lemma 3.4, and rescale the time coordinate by $t=4m\tau$. Then 
\be \frac{dv}{d\tau} = (1+\lambda)v-v^2,\quad \tau\geq\tau_0.\label{eq:dv}\ee
Note that $\tau\mapsto\lambda(\tau)$ is continuous on $[\tau_0,+\infty)$ and satisfies $\lim_{\tau\to+\infty}\lambda(\tau)=0$ (and $\tau_0=t_0/4m$). We complete the proof by establishing that 
\be \lim_{\tau\to+\infty} v(\tau) =1. \label{eq:v-lim} \ee
We can exploit the Ricatti form of (\ref{eq:dv}) to write the solution as 
\be v(\tau) = \frac{v_0 k(\tau)}{1+v_0\int_{\tau_0}^\tau k(\tau')d\tau'},\quad v_0=v(\tau_0)\in (0,1) \label{eq:v-sol} \ee
where 
\be k(\tau) = e^{\tau-\tau_0}\exp\int_{\tau_0}^\tau \lambda(\tau')d\tau'.\label{eq:k-intdef} \ee
Integrating by parts yields 
\be \int_{\tau_0}^\tau k(\tau')d\tau'=k(\tau)-1 -\int_{\tau_0}^\tau \lambda(\tau')k(\tau')d\tau',\label{eq:k-int-parts}\ee
so that 
\be \int_{\tau_0}^\tau (1+\lambda(\tau'))k(\tau')d\tau'=k(\tau)-1.\label{eq:k-int-rearrange} \ee
We approximate the integral on the left by applying the generalised mean value theorem for integrals and conclude that for every $\tau>\tau_0$, there exists $\tau_*\in(\tau_0,\tau)$ such that 
\be \int_{\tau_0}^\tau k(\tau')d\tau' = \frac{k(\tau)-1}{1+\lambda(\tau_*)}, \label{eq:k-int-approx}\ee
and so 
\be v(\tau) = \frac{v_0(1+\lambda(\tau_*))k(\tau)}{1+\lambda(\tau_*)-v_0+v_0k(\tau)}.\label{eq:v-star} \ee
By (\ref{eq:lam-lim}) and positivity of $v_0$ and $k$, we see that the numerator here is positive for all sufficiently large $\tau_0$. Since we also know that $v(\tau)>0$, the denominator must also be positive. 
Then 
\begin{eqnarray}
    |1-v| &=& \left| \frac{(1+\lambda(\tau_*))(1+\lambda(\tau_*)-v_0)}{1+\lambda(\tau_*)-v_0+v_0k(\tau)}-\lambda(\tau_*)\right| \nonumber \\
    &\leq & \frac{|(1+\lambda(\tau_*))(1+\lambda(\tau_*)-v_0)|}{1+\lambda(\tau_*)-v_0+v_0k(\tau)} + |\lambda(\tau_*)|. \label{eq:v-bound} 
\end{eqnarray}
Using (\ref{eq:lam-lim}) again, we see that for any $\epsilon>0$, there exists $\tau_1$ such that 
\be -\frac{\epsilon}{2}<\lambda(\tau)<\frac{\epsilon}{2},\quad \tau>\tau_1,\label{eq:lam-bounds} \ee
and so from(\ref{eq:k-def}) 
\be k(\tau) > e^{(1-\epsilon/2)(\tau-\tau_0)},\quad \tau>\tau_1. \label{eq:k-lower} \ee
It follows that both the first and second terms on the right hand side of (\ref{eq:v-bound}) can be made arbitrarily small for all sufficiently large $\tau$, and (\ref{eq:v-lim}) follows. \hfill$\blacksquare$

\subsubsection{Curvature scalars at the horizon}

These last three lemmas provide the key information required to determine the behaviour of the zero-, first- and second-order curvature invariants mentioned above. We do this by imposing the asymptotics of Lemma 3.3 and combining terms to identify polynomials in $K$. The coefficients of these will yield the required limits. For example, we have 

\be R_{abcd}R^{abcd} \sim 48\frac{m^2}{r^6}+24H^4 +24\kappa H^3K+12\kappa^2H^2K^2,\quad t\to+\infty,\label{eq:riem0-lim1} \ee

giving 

\be \lim_{t\to\infty} R_{abcd}R^{abcd} = \frac{3}{4m^4}.\label{eq:riem0-lim} \ee

Applying the same prescription to the first order term (\ref{eq:kretsch1}) gives 
\begin{eqnarray}
 \nabla_eR_{abcd}\nabla^eR^{abcd} &\sim& 720\frac{m^2}{r^8}\alpha+(28\kappa^2\frac{m^2}{r^4}-48\kappa^4H^2)K^4\nonumber \\
 &&  +48\kappa^3\left(2-\frac{5m}{r}\right)HK^5 \nonumber \\  && -12\kappa^2\left(12-52\frac{m}{r}+57\frac{m^2}{r^2}\right)K^6,\quad t\to+\infty \label{eq:riem1-lim1} 
\end{eqnarray}

and so

\be \lim_{t\to+\infty} \nabla_eR_{abcd}\nabla^eR^{abcd} = \frac{\kappa^2}{16m^6}. \label{eq:riem1-lim} \ee

We note a slightly curious aspect to these results. In each case, the first term on the right hand side of (\ref{eq:riem0-lim1}) and (\ref{eq:riem1-lim1}) is identically the contribution from the Weyl tensor. This term has a finite, non-zero limit at zero order, but a vanishing limit at first order. So derivatives do not necessarily introduce a more strongly divergent quantity.

Repeating this process for the second order term, we find 
\be \nabla_f\nabla_eR_{abcd}\nabla^f\nabla^eR^{abcd} \sim -\frac{4}{r^{17}}\sum_{i=0}^4 c_{2i}(r) H^{2i},\quad t\to+\infty. \label{eq:riem2-lim1} \ee
The coefficients $c_{2i},i=0,\dots,4$ are given in an appendix. 
Since $K=H\gamma^{1/2}\to1/2m$ (with $H\to0,\gamma\to+\infty$) in the limit, we see that a term with $\gamma^k$ in $c_{2i}$ is infinite (respectively, non-zero and finite, zero) in the limit if and only if $k>i$ (respectively, $k=i, k<i$). 

Collecting the relevant coefficients - which come from the highest powers of $\gamma$ in $c_4,c_6$ and $c_8$ - we find that 
\be \nabla_f\nabla_eR_{abcd}\nabla^f\nabla^eR^{abcd} \sim \frac{25}{512}\frac{\kappa^2}{m^8}\gamma,\quad t\to+\infty, \label{eq:riem2-lim2} \ee
and so 
\be 
\lim_{t\to+\infty} \nabla_f\nabla_eR_{abcd}\nabla^f\nabla^eR^{abcd} =+\infty. 
\label{eq:riem2-lim} \ee

This establishes rigorously the claim of \cite{kaloper2010mcvittie} that this term diverges at the horizon, and extends the result from radial null geodesics to all causal geodesics that meet the horizon. (We note that \cite{lake2011more} found a finite limit for this quantity.) We collect the results above as follows:

\begin{proposition}
     Consider an expanding McVittie spacetime with a Big Bang background satisfying (\ref{eq:H-cons})-(\ref{eq:scale-fac-lim}) with $H_0=\Lambda=0$. Assume further that the background FLRW spacetime has an equation of state such that the asymptotic relations (\ref{eq:Hp-lim})-(\ref{eq:Hppp-lim}) hold.  
     Then as the black hole horizon is approached along a causal geodesic $C$, the zero-order curvature invariants $\Psi_2, R, R_{ab}R^{ab}$ and $R_{abcd}R^{abcd}$ and the first order curvature invariant $\nabla_eR_{abcd}\nabla^eR^{abcd}$ have finite limits along $C$. The second order term $\nabla_f\nabla_eR_{abcd}\nabla^f\nabla^eR^{abcd}$ diverges. \hfill$\blacksquare$
\end{proposition}

\section{Jacobi fields and tidal forces at the horizon.}

For black holes in vacuum, all curvature terms are finite at the horizon, and so no pathological behaviour (e.g.\ destructive tidal forces) is encountered on crossing the horizon. It is less clear that this is the case in a McVittie spacetime, due to the singular behaviour of the pressure when $r\to 2m$ on surfaces of constant $t$. (As seen above, there is no issue if $\Lambda>0$, and so the the discussion here applies to the case $\Lambda=0$.) But Proposition 3.1 above shows that there is no divergence of zero or first order curvature terms along timelike geodesics crossing the horizon. These results relate to the behaviour of scalar curvature terms, and as we now show, also carry over to tidal forces - at least in the case of radial timelike geodesics. Consider three independent solutions of the geodesic deviation equation 
\be D^2X^a = {R_{bcd}}^av^bv^dX^b \label{eq:GDE} \ee 
along a radial timelike geodesic $C$ with tangent $v^a$ - that is, we consider three independent Jacobi fields $\{X_i, i=1,2,3\}$. The results quoted here on Jacobi fields in spherical symmetry are from \cite{nolan1999strengths}. We can take $X_1, X_2$ to be tangent to the 2-spheres of symmetry lying in the span of $\{\partial_\theta,\partial_\phi\}$: the norm of each is a constant multiple of 
\be X(s) = \int_{s_i}^s \frac{du}{r^2(u)} \label{eq:jnorm1} \ee
where $s_i$ is some initial value of the proper time on $C$. 
 In any spherically symmetric spacetime, the norm $W$ of the third Jacobi field (orthogonal to the 2-spheres, and of course to the tangent to the geodesic) satisfies 
\be \ddot{W} + 2(\frac{E}{r^3}+2\Psi_2-\frac{R}{12})W = 0,\label{eq:jnorm2} \ee
where
\be E = \frac{r}{2}\left(1-g^{ab}\nabla_ar\nabla_br\right) \label{eq:MSE} \ee 
is the Misner-Sharp mass \cite{misner1964relativistic} (or alternatively the gravitational energy in spherical symmetry: see \cite{hayward1996gravitational}), $\Psi_2$ is the Weyl scalar as above, and $R$ is the Ricci scalar. The overdots indicate derivatives with respect to proper time along the geodesic.  In McVittie's spacetime, (\ref{eq:jnorm2}) reads
\be \ddot{W} - (\frac{2m}{r^3}+H^2+H'\gamma^{1/2})W = 0. \label{eq:jnorm3} \ee
It follows from Lemmas 3.3 and 3.4 that the coefficient of $W$ has a finite limit as the geodesic crosses the horizon, and so the general solution of (\ref{eq:jnorm3}) is finite and non-zero. Thus in the language of \cite{ori2000strength}, the singularity along the black hole horizon is not deformationally strong (nor is it strong in Tipler's sense \cite{tipler1977singularities}). 

However, this is not the whole story. Evidently, those parts of the curvature felt by Jacobi fields along a radial timelike geodesic miss other curvature components which are infinite in the relevant limit. We can show that there are parallel propagated components of the curvature tensor that are infinite in the limit as the horizon is approached (Proposition 4.1 below). 

Let $\{e_i,i=0,1,2,3\}$ be a pseudo-orthonormal frame of vectors parallel propagated along a radial timelike geodesic $C$ with tangent $v=e_0$, so that 
\be \nabla_v e_i=0,\quad g(e_i,e_j)=\eta_{ij}=\textrm{diag}(-1,1,1,1).\label{eq:pp-frame} \ee
Then the parallel propagated frame components of the curvature tensor along $C$ are given by 
\be \mathrm{R}_{ijkl}=R_{abcd}e_i^ae_j^be_k^ce_l^d. \label{eq:Riem-frame} \ee
Using the Ricci decomposition (\ref{eq:riem-decomp}) and the field equations, we can write
\be R_{abcd} = C_{abcd}+2H^2g_{a[c}g_{d]b}-2H'\gamma^{1/2}(g_{a[c}u_{d]}u_b-g_{b[c}u_{d]}u_a). \label{eq:riem-udecomp} \ee
In the coordinates $x^\alpha=(t,r,\theta,\phi)$, we have $v=e_0=\dot{t}\partial_t+\dot{r}\partial_r$ and we can take 
\be e_2 = \frac{1}{r}\partial_\theta,\quad e_3=\frac{1}{r\sin\theta}\partial_\phi.\label{eq:e23} \ee
A direct calculation using Mathematica gives (with hopefully obvious notation) 
\be \mathrm{C}_{0i0j} = \frac{m}{r^3}\delta_{ij},\quad i,j\in\{2,3\}, \label{eq:weyl-frame-terms} \ee
and so (recalling that $u$ is given by (\ref{u-mcv}))
\be \mathrm{R}_{0i0j} = \left(\frac{m}{r^3}-H^2-H'\gamma^{-1/2}\dot{t}^2\right)\delta_{ij},\quad i,j\in\{2,3\}.\label{eq:r0i0j} \ee
The key term here is the last one, and using Lemmas 3.3 and 3.4 we have
\be H'\gamma^{-1/2}\dot{t}^2\sim -2m\kappa H^3\dot{t}^2,\quad t\to+\infty.\label{eq:key} \ee
From the proof of Theorem 2.1, we know that $u=\dot{t}\to+\infty$ as $t\to+\infty$ along a causal geodesic running into the black hole horizon. Thus to determine the limiting behaviour of the tidal force term $\mathrm{R}_{0i0j}$ at the horizon, we must determine the relative rates of $H$ and $\dot{t}$. We argue as follows that this term diverges in the limit $t\to\infty$. As $\dot{r}<0$ for sufficiently large $t$, we have from (\ref{eq:geo-t2}) 
\be \ddot{t} > -\left(1-\frac{3m}{r}\right)H\gamma^{1/2}\dot{t}^2 \sim \frac{1}{4m}\dot{t}^2. \label{eq:tdd-asymp} \ee
Integrating shows that $\dot{t}$ diverges exponentially in $t$ as $t\to+\infty$. But integrating (\ref{eq:Hp-lim}) yields $H\sim \frac{1}{\kappa t}$. It follows that $H^3\dot{t}^2$ diverges in the limit as $t\to+\infty$ along the geodesic. 

\begin{proposition}
The parallel propagated frame components $\mathrm{R}_{0i0i}, i=2,3$ diverge in the limit as the black hole horizon is approached along a radial timelike geodesic. \hfill$\blacksquare$     
\end{proposition}
    
\section{Conclusions} 

The main result of this paper (Theorem 2.1) fills an important gap in the interpretation of McVittie spacetimes. As described above, the black hole nature of the spacetime was first established in \cite{kaloper2010mcvittie}. This and subsequent papers dealt with the behaviour of null geodesics of the spacetime (mainly radial null geodesics), allowing one to connect with the classical definition of a black hole of (for example) \cite{hawking1973large} or \cite{wald1984general}. Particle orbits of McVittie spacetimes have been studied on a number of occasions, often in relation to the question of how such trajectories are influenced by the expansion of the universe. McVittie himself initiated this line of work \cite{mcvittie1933mass}, and see also \cite{noerdlinger1971effect, carrera2010influence, nolan2014particle, jayswal2025local} and further references in the last of these. However, the question addressed here seems not to have been considered before, and has a clear answer: yes, you can fall into a McVittie black hole. 

The question of what happens along the way is less straightforward - at least in the case $H_0=\Lambda=0$. In the case of a positive cosmological constant, the answer seems clear: nothing much happens, and an observer can fall through the black hole horizon into the interior of a Schwarzschild-de Sitter black hole. With a vanishing cosmological constant, as we have seen, zero- and first-order curvature scalars remain finite, but the second order term (\ref{eq:riem2-lim}) diverges along any causal geodesic running into the black hole horizon (as first noted in \cite{kaloper2010mcvittie} in the case of radial null geodesics). Jacobi fields (deviation vectors) along a radial timelike geodesic remain finite, but at least some tidal forces diverge as per Proposition 4.1. There is no contradiction involved in this last point: the divergent tidal forces are integrable, yielding finite solutions of the geodesic deviation equation. In this scenario, and more generally when a geodesic carries along with it finite Jacobi fields (regardless of a potential divergence of tidal forces), it has been argued that this indicates that a small test body - modelled by an orthogonal triad of Jacobi fields along the geodesic - can survive the impact with the singularity: the singularity is gravitationally (or deformationally) weak \cite{ori2000strength,tipler1977singularities}. We argue that there are reasons to question this interpretation in cases such as the present one, where higher order curvature terms diverge. 

First, we note that there are different approaches to the geodesic deviation equation (GDE) as discussed in \cite{vines2015geodesic}. These different approaches in fact involve different definitions of the deviation vector that connects different members of a congruence of geodesics. In \cite{wald1984general}, this is taken to be the variation with respect to the parametrization of a one-parameter family of geodesics, Lie transported along the geodesic. This yields the exact GDE (\ref{eq:GDE}). In other approaches, the deviation vector is taken to be the tangent to the unique spacelike vector connecting two members of the congruence at equal proper times. See \cite{vines2015geodesic} for details. This can be described in the language of bitensors and Synge's world function $\sigma$, so that the tangent is $\xi^\alpha=\nabla^\alpha\sigma(x,x')$, with $x,x'$ being points on different members of the congruence. In this treatment, the GDE provides a first order approximation to the evolution of $\xi^\alpha$ along members of the congruence. Crucially for our purposes, \textit{higher order approximations to the equation of motion for $\xi^\alpha$ involve higher order derivatives of the curvature}.  So in this approach, the GDE (\ref{eq:GDE}) provides only a linear approximation, and caution must be applied in interpreting the physical significance of having finite solutions of this approximation - especially when we know that higher order approximations will bring in the divergent second order curvature terms. 

Second, we consider another (and more realistic) description of extended bodies in General Relativity, namely the framework developed by Dixon (see \cite{dixon1970dynamicsI,dixon1970dynamicsII,dixon1974dynamicsIII} and \cite{harte2015motion}). In this framework the extended body is characterized by an infinite set of multipole moments defined by integrals of the energy-stress-moment tensor. The quadrupole and higher moments generate force terms that arise in the equations of motion of the momentum and spin of the body. In these force terms, the $2^n-$pole moment couples to derivatives of the Riemann tensor of order $n-1$. The approach typically involves a finite cut-off (i.e.\ a finite number of multipole moments are taken into account), but we see that including the octupole moment will bring in the second order derivative of the curvature tensor, which as we have seen includes divergences. While we will not pursue the relevant calculations here, it would seem incautious to make a definitive statement about the survival of an extended body falling into a black hole without knowing how these divergent terms contribute to its motion. These calculations would be highly involved: a more fruitful way of determining the physical effects of the singular aspects of the horizon may be to consider (for example) the behaviour of a test scalar field or test fluid that impinges on the horizon. A study of the latter has yielded interesting results in relation to fluids impacting on weak null singularities in black hole spacetimes \cite{mancheva2025}.



\section*{Acknowledgements}
Thanks to Abraham Harte, Alex Grant, Tomohiro Harada, Raya Mancheva and Jos\'e Senovilla for useful conversations. Thanks also to an anonymous referee for helpful comments on the original version of this paper.

\appendix 

\section{Second order curvature scalars}
For the second-order term (\ref{eq:kretsch2}), we find
\begin{eqnarray}
    D2\textrm{Weyl} &=&
    1440\frac{m^2}{r^{10}}\left(14-60\frac{m}{r}+65\frac{m^2}{r^2}\right)+2880\frac{m^2}{r^8}\left(12-25\frac{m}{r}\right)H^2 \nonumber \\
    && + 2^5495\frac{m^2}{r^6}H^4-1440\frac{m^3}{r^9}\gamma^{1/2}H'\nonumber \\
    && -4320\frac{m^2}{r^6}\gamma^{1/2}H^2H'+720\frac{m^2}{r^6}\gamma(H')^2\label{eq:ddweyl-squared} 
\end{eqnarray}
and
\begin{eqnarray} 
D2\textrm{Ric} &=& 12\gamma^3(H^{(3)})^2 + 24\left(2-\frac{7m}{r}\right)\gamma^{7/2}HH''H^{(3)}+56\frac{m^2}{r^4}\gamma^3H'H^{(3)} \nonumber \\
&& +24\left(2-\frac{5m}{r}\right)(H')^2H^{(3)}-24\frac{m}{r}\left(1-\frac{5m}{r}\right)\gamma^4H^2H'H^{(3)} \nonumber \\
&& - 160\frac{m^2}{r^4}\gamma^3(H'')^2 + 12\left(19-88\frac{m}{r}+109\frac{m^2}{r^2}\right)\gamma^4H^2(H'')^2\nonumber \\
&&+24 \frac{m^2}{r^4}\left(8-\frac{3m}{r}\right)\gamma^{7/2}HH'H''\nonumber \\
&& -24\left(2+5\frac{m}{r}-53\frac{m^2}{r^2}+79\frac{m^3}{r^3}\right)\gamma^{9/2}H^3H'H''\nonumber \\
&& + 24\left(12-56\frac{m}{r}+67\frac{m^2}{r^2}\right)\gamma^4H(H')^2H''\nonumber \\
&& - 4\frac{m^2}{r^6}\left(42-112\frac{m}{r}+127\frac{m^2}{r^2}\right)\gamma^3(H')^2\nonumber \\
&& -8\frac{m^2}{r^4}\left(96-347\frac{m}{r}+334\frac{m^2}{r^2}\right)\gamma^4H^2(H')^2 \nonumber \\
&& +12\left(52-412\frac{m}{r}+1240\frac{m^2}{r^2}-1686\frac{m^3}{r^3}+885\frac{m^4}{r^4}\right)\gamma^5H^4(H')^2 \nonumber \\
&&-8\frac{m^2}{r^4}\left(2+\frac{3m}{r}\right)\gamma^{7/2}(H')^3\nonumber \\ &&-24\frac{m}{r}\left(10-47\frac{m}{r}+57\frac{m^2}{r^2}\right)\gamma^{9/2}H^2(H')^3\nonumber \\
&&+12\left(12-52\frac{m}{r}+57\frac{m^2}{r^2}\right)\gamma^4(H')^4.
\label{eq:D2Ric-squared}
\end{eqnarray}
We note that in (\ref{eq:ddweyl-squared})-(\ref{eq:D2Ric-squared}), none of the polynomials of degree one, two and three in $m/r$ has a root at $r=2m$. 

\section{Coefficients of $\nabla_f\nabla_eR_{abcd}\nabla^f\nabla^eR^{abcd}$}
The coefficients in (\ref{eq:riem2-lim1}) are given below. These are written as sums of descending positive powers of $\gamma^{1/2}$. The coefficients are non-vanishing at $r=2m$.
\begin{eqnarray}
    c_0 &=& -360m^2r^7\left(14-60\frac{m}{r}+65\frac{m^2}{r^2}\right),\\
    c_2 &=& -360\kappa m^3r^8\gamma^{1/2}+720m^2r^9\left(12-25\frac{m}{r}\right),\\
    c_4 &=& -\kappa^2m^2r^{11}\left(222-832\frac{m}{r}+847\frac{m^2}{r^2}\right)\gamma^3-1080\kappa m^2r^{11}\gamma^{1/2}\nonumber \\
    && -3960m^2r^{11},\\
    c_6 &=& 2\kappa^2m^2r^{13}\left(96-347\frac{m}{r}+334\frac{m^2}{r^2}\right)\gamma^4 + 2\kappa^3 m^2 r^{13}\left(46-21\frac{m}{r}\right)\gamma^{7/2} \nonumber \\
    && + 104 \kappa^4m^2r^{13}\gamma^3,\\
    c_8&=& -3\kappa^2r^{17}\left(52-412\frac{m}{r}+1240\frac{m^2}{r^2}-1686\frac{m^3}{r^3}+885\frac{m^4}{r^4}\right)\gamma^5\nonumber\\
    && -6\kappa^3r^{17}\left(4+20\frac{m}{r}-153\frac{m^2}{r^2}+215\frac{m^3}{r^3}\right)\gamma^{9/2} \nonumber \\
    &&-3\kappa^4r^{17}\left(136-636\frac{m}{r}+801\frac{m^2}{r^2}\right)\gamma^4 \nonumber \\
    && +24\kappa^5r^{17}\left(6-19\frac{m}{r}\right)\gamma^{7/2}-48\kappa^6r^{17}\gamma^3. 
\end{eqnarray}

\section*{References}
\bibliographystyle{unsrt}
\bibliography{Falling_McV_BH_R1}

\begin{thebibliography}{10}

\bibitem{mcvittie1933mass}
G.C. McVittie.
\newblock {The mass-particle in an expanding universe}.
\newblock {\em Monthly Notices of the Royal Astronomical Society}, 93:325, 1933.

\bibitem{sussman1988spherically}
R.A. Sussman.
\newblock {On spherically symmetric shear-free perfect fluid configurations (neutral and charged). III. Global view}.
\newblock {\em Journal of Mathematical Physics}, 29:1177, 1988.

\bibitem{kaloper2010mcvittie}
N.~Kaloper, M.~Kleban, and D.~Martin.
\newblock {McVittie's legacy: black holes in an expanding universe}.
\newblock {\em Physical Review D}, 81:104044, 2010.

\bibitem{lake2011more}
K.~Lake and M.~Abdelqader.
\newblock {More on McVittie's legacy: A Schwarzschild--de Sitter black and white hole embedded in an asymptotically $\Lambda$ CDM cosmology}.
\newblock {\em Physical Review D}, 84:044045, 2011.

\bibitem{nolan2017local}
B.C. Nolan.
\newblock {Local properties and global structure of McVittie spacetimes with non-flat Friedmann--Lema{\^\i}tre--Robertson--Walker backgrounds}.
\newblock {\em Classical and Quantum Gravity}, 34(22):225002, 2017.

\bibitem{Poplawski_2025}
Nikodem Popławski.
\newblock Black holes in the expanding universe.
\newblock {\em Classical and Quantum Gravity}, 42(6):065017, 2025.

\bibitem{wald1984general}
RM~Wald.
\newblock {\em General relativity}.
\newblock Chicago, University of Chicago Press, 1984.

\bibitem{nolan1999point}
B.C. Nolan.
\newblock {A point mass in an isotropic universe: II. Global properties}.
\newblock {\em Classical and Quantum Gravity}, 16:1227, 1999.

\bibitem{ellis2012relativistic}
George~FR Ellis, Roy Maartens, and Malcolm~AH MacCallum.
\newblock {\em Relativistic cosmology}.
\newblock Cambridge University Press, 2012.

\bibitem{nolan2014particle}
B.C. Nolan.
\newblock {Particle and photon orbits in McVittie spacetimes}.
\newblock {\em Classical and Quantum Gravity}, 31(23):235008, 2014.

\bibitem{perkodifferential}
L.~Perko.
\newblock {\em {Differential equations and dynamical systems}}.
\newblock Springer-Verlag, New York, 1991.

\bibitem{nolan1999strengths}
Brien~C Nolan.
\newblock Strengths of singularities in spherical symmetry.
\newblock {\em Physical Review D}, 60(2):024014, 1999.

\bibitem{misner1964relativistic}
Charles~W Misner and David~H Sharp.
\newblock Relativistic equations for adiabatic, spherically symmetric gravitational collapse.
\newblock {\em Physical Review}, 136(2B):B571, 1964.

\bibitem{hayward1996gravitational}
Sean~A Hayward.
\newblock Gravitational energy in spherical symmetry.
\newblock {\em Physical Review D}, 53(4):1938, 1996.

\bibitem{ori2000strength}
Amos Ori.
\newblock Strength of curvature singularities.
\newblock {\em Physical Review D}, 61(6):064016, 2000.

\bibitem{tipler1977singularities}
Frank~J Tipler.
\newblock Singularities in conformally flat spacetimes.
\newblock {\em Physics Letters A}, 64(1):8--10, 1977.

\bibitem{hawking1973large}
S.W. Hawking and G.F.R. Ellis.
\newblock {\em {The large scale structure of space-time}}.
\newblock Cambridge University Press, 1973.

\bibitem{noerdlinger1971effect}
Peter~D Noerdlinger and Vah{\'e} Petrosian.
\newblock The effect of cosmological expansion on self-gravitating ensembles of particles.
\newblock {\em Astrophysical Journal, vol. 168, p. 1}, 168:1, 1971.

\bibitem{carrera2010influence}
Matteo Carrera and Domenico Giulini.
\newblock Influence of global cosmological expansion on local dynamics and kinematics.
\newblock {\em Reviews of Modern Physics}, 82(1):169--208, 2010.

\bibitem{jayswal2025local}
Vishal Jayswal and Sergei~M Kopeikin.
\newblock Local coordinates and motion of a test particle in the mcvittie spacetime.
\newblock {\em Classical and Quantum Gravity}, 42(6):065023, 2025.

\bibitem{vines2015geodesic}
Justin Vines.
\newblock Geodesic deviation at higher orders via covariant bitensors.
\newblock {\em General Relativity and Gravitation}, 47(5):59, 2015.

\bibitem{dixon1970dynamicsI}
William~G Dixon.
\newblock Dynamics of extended bodies in general relativity. i. momentum and angular momentum.
\newblock {\em Proceedings of the Royal Society of London. A. Mathematical and Physical Sciences}, 314(1519):499--527, 1970.

\bibitem{dixon1970dynamicsII}
William~G Dixon.
\newblock Dynamics of extended bodies in general relativity-ii. moments of the charge-current vector.
\newblock {\em Proceedings of the Royal Society of London. A. Mathematical and Physical Sciences}, 319(1539):509--547, 1970.

\bibitem{dixon1974dynamicsIII}
William~G Dixon.
\newblock Dynamics of extended bodies in general relativity iii. equations of motion.
\newblock {\em Philosophical Transactions of the Royal Society of London. Series A, Mathematical and Physical Sciences}, 277(1264):59--119, 1974.

\bibitem{harte2015motion}
Abraham~I Harte.
\newblock Motion in classical field theories and the foundations of the self-force problem.
\newblock In {\em Equations of Motion in Relativistic Gravity}, pages 327--398. Springer, 2015.

\bibitem{mancheva2025}
Raya Mancheva.
\newblock The contrasting behaviour of two perfect fluids near the weak null singularity of a spherically symmetric black hole.
\newblock Poster presentation at the 24th International Conference on General Relativity and Gravitation, 2025.

\end{thebibliography}

\end{document}